\documentclass[conference]{IEEEtran}
\IEEEoverridecommandlockouts
\usepackage{cite}
\usepackage{amsmath,amssymb,amsfonts}
\usepackage{algorithmic}
\usepackage{graphicx}
\usepackage{textcomp}
\usepackage{xcolor}
\usepackage{comment}
\def\BibTeX{{\rm B\kern-.05em{\sc i\kern-.025em b}\kern-.08em
    T\kern-.1667em\lower.7ex\hbox{E}\kern-.125emX}}
\begin{document}

\title{In-memory multiplication engine with SOT-MRAM based stochastic computing\\
}

\author{Xin Ma$^{1,2}$, Liang Chang$^{1,3}$, Shuangchen Li$^{1}$, Lei Deng$^{1}$, Yufei Ding$^{2}$, Yuan Xie$^{1}$\\
\\
$^1$Department of Electrical and Computer Engineering, UCSB, California, USA\\
$^2$Department of Computer Science, UCSB, California, USA\\
$^3$School of Electronic and Information Engineering, Behang University, Beijing, China}

\maketitle

\begin{abstract}
Processing-in-memory (PIM) turns out to be a promising solution to breakthrough the memory wall and the power wall. While prior PIM designs yield successful implementation of bitwise Boolean logic operations locally in memory, it is difficult to accomplish the multiplication (MUL) instruction in a fast and efficient manner. In this paper, we propose a new stochastic computing (SC) design to perform MUL with in-memory operations. Instead of using the stochastic number generators (SNGs), we harness the inherent stochasticity in the memory write behavior of the magnetic random access memory (MRAM). Each memory bit serves as an SC engine, performs MUL on operands in the form of write voltage pulses, and stores the MUL outcome in-situ. The proposed design provides up to 4x improvement in performance compared with conversational SC approaches, and achieves 18x speedup over implementing MUL with only in-memory bitwise Boolean logic operations. 

\end{abstract}


\section{Introduction}
The processing-in-memory (PIM) paradigm has been considered as a promising alternative to break the bottlenecks of conventional von-Neumann architecture. In the era of big data, data movement between the processor and the memory results in huge power consumption (power wall) and performance degradation (memory wall), known as the von-Neumann bottleneck\cite{koo2017summarizer}. By placing the processing units inside/near the memory, PIM remarkably reduces the energy and performance overhead induced by data transport\cite{zhang2014top}\cite{ahn2016scalable}. In the recent advancement of PIM designs, it also allows fully leverage the large memory internal bandwidth and embrace massive parallelism by simultaneously activating multiple rows, subarrays, and banks in memory arrays for bit-wise operations \cite{li2016pinatubo}. These performance gains are all achieved at a minimal cost of slightly modifying the memory array peripheral circuits \cite{ chi2016prime} \cite{li2017drisa}. 

Multiplication (MUL) is always a complex task to accomplish in efficient PIM designs, despite that MUL instructions are frequently used in Neuro Network (NN) algorithms and linear transforms (e.g. Discrete Fourier Transform). As shown by the recent developed DRISA \cite{li2017drisa}, it takes 143 cycles to calculate an 8-bit multiplication, which deviates from its original motivation to achieve high performance with in-memory bitwise operations. The situation may be even worse with operands composing of more bits, as the cycle count can increase exponentially with the operand's bit length. The challenge mainly lies in the fact that MUL can not be effectively decomposed into a small serial of bitwise Boolean logic operations which can be performed locally in memory. 

To tackle such challenge, prior efforts propose to either approximate the MUL or utilize the analog computing features of hardware devices. On one hand at the algorithm level, binary NN with approximate binary weights and activations has been developed \cite{courbariaux2016binarized}. As such, the MUL is simplified into bitwise XNOR operations that become PIM friendly \cite{angizi2018imce}. Unfortunately, such simplification comes at the cost of the undesired and significant degradation in the classification accuracy of NN. On the other hand at the hardware level, ReRAM is implemented to ease MUL in novel PIM designs, taking advantage of ReRAM's analog storage. The analog resistance/conductance of ReRAM encodes the weights in NN. By activating one entire row/column simultaneously in a ReRAM crossbar, the dot product between a matrix and a vector in NN can be easily achieved using Ohm's law\cite{ chi2016prime}. Nevertheless, ReRAM itself suffers from the long write latency, high programming voltage and limited endurance, which hinders its application in high-speed and energy efficient architecture design. 

In this work, we propose a new stochastic computing (SC) design to effectively perform MUL with in-memory operations, in light of the simplicity to implement MUL with SC. In order to tightly couple SC with PIM, we embrace the inherent stochasticity of the memory bit in spin-orbit-torque magnetic random access memory (SOT-MRAM). Specifically, the stochastic number generation and massive AND operations in the conventional SC-based MUL are implemented with simple memory write operations in SOT-MRAM. Consequently, each bit serves as an SC engine, and the large supporting circuits for stochastic number generation and logic operations can be effectively saved. Finally, the MUL outcome is represented by the probability distribution of the binary storage states among MRAM bits, and can be converted back to its binary form with pop-count. The contributions of this paper are summarized as follows:
\begin{itemize}
\item We propose the idea of employing the inherent stochastic write in SOT-MRAM to promote SC in the PIM design. 
\item We develop an efficient approach to implement MUL in the way of memory write, by converting the binary multipliers to the write voltage pulse with varied duration.
\item We propose two strategies of pop-count to convert the MUL result back to its binary format, offering flexibility to further trade performance with area.
\item The proposed design provides up to 4x improvement in performance and significant reduction in area occupancy compared with conversational SC approaches, and achieves 18x speedup over implementing MUL with only in-memory bitwise Boolean logic operations. 
\end{itemize}

\section{Preliminaries}
This section introduces the motivation to combine SC with PIM and the preliminary design with the stochastic switching behavior of SOT-MRAM.

\subsection{SC and PIM}

SC provides an alternative approach to implement the MUL function. SC is an approximate computing method, which has been studied for decades and widely applied to image/signal processing, control systems and general purpose computing\cite{hayes2015introduction}\cite{alaghi2013survey}. SC method essentially trades the data representation density for simpler logic design and lower power. For instance, SC represents a n-bit binary number with a stochastic bitstream (~$2^n$-bit). The value of the binary number $X$ equals to the probability of the appearance of "1"s in the bitstream $(\{x_i\})$:
\begin{equation}
X=2/6(binary)\rightarrow \{x_i\}=\{1,0,0,0,1,0\}(stoch.),
\end{equation}
Benefiting from such data representation, the MUL between two numbers can be converted to simple bitwise AND operations, which dramatically reduces the complexity of logic design. 
\begin{equation}
\begin{split}
&Y=3/6(binary)\rightarrow \{x_i\}=\{0,1,0,1,1,0\}(stoch.),\\
&X \bullet Y=1/6 \rightarrow \{x_i\&y_i\}=\{0,0,0,0,1,0\}(stoch.).
\end{split}
\end{equation}
However, SC is not friendly to conventional von-Neumann architecture. The data explosion of SC aggravates data movement between processor and memory, which offsets the simplicity brought by SC.   

Instead, SC tightly couples with PIM from multi-fold aspects, leading to significant performance gain: First, the many bits of stochastic bitstream can be stored in off-chip memory with large capacity. Second, the logic operation with reduced complexity can be implemented by the processing units locally in memory. Finally, the stochastic feature of bitstream allows parallel computing on the individual bit, so that the internal memory bandwidth can be fully leveraged. Therefore, the MUL instruction can be significantly accelerated by combining SC with PIM. 

Several challenges still exist towards combine SC with PIM. The random bitstream still relies on stochastic number generators (SNGs), which incurs large area overhead for the supporting circuits. In addition, those stochastic bits can be hardly generated in parallel and with eliminated correlations, resulting in degradation of performance and accuracy in computing MUL. In our design, we overcome these drawbacks by utilizing the inherent stochasticity in MRAM bit.

\subsection{SOT-MRAM and its stochastic switching}

\begin{figure}[htbp]
\centering
\centerline{\includegraphics[width=3 in] {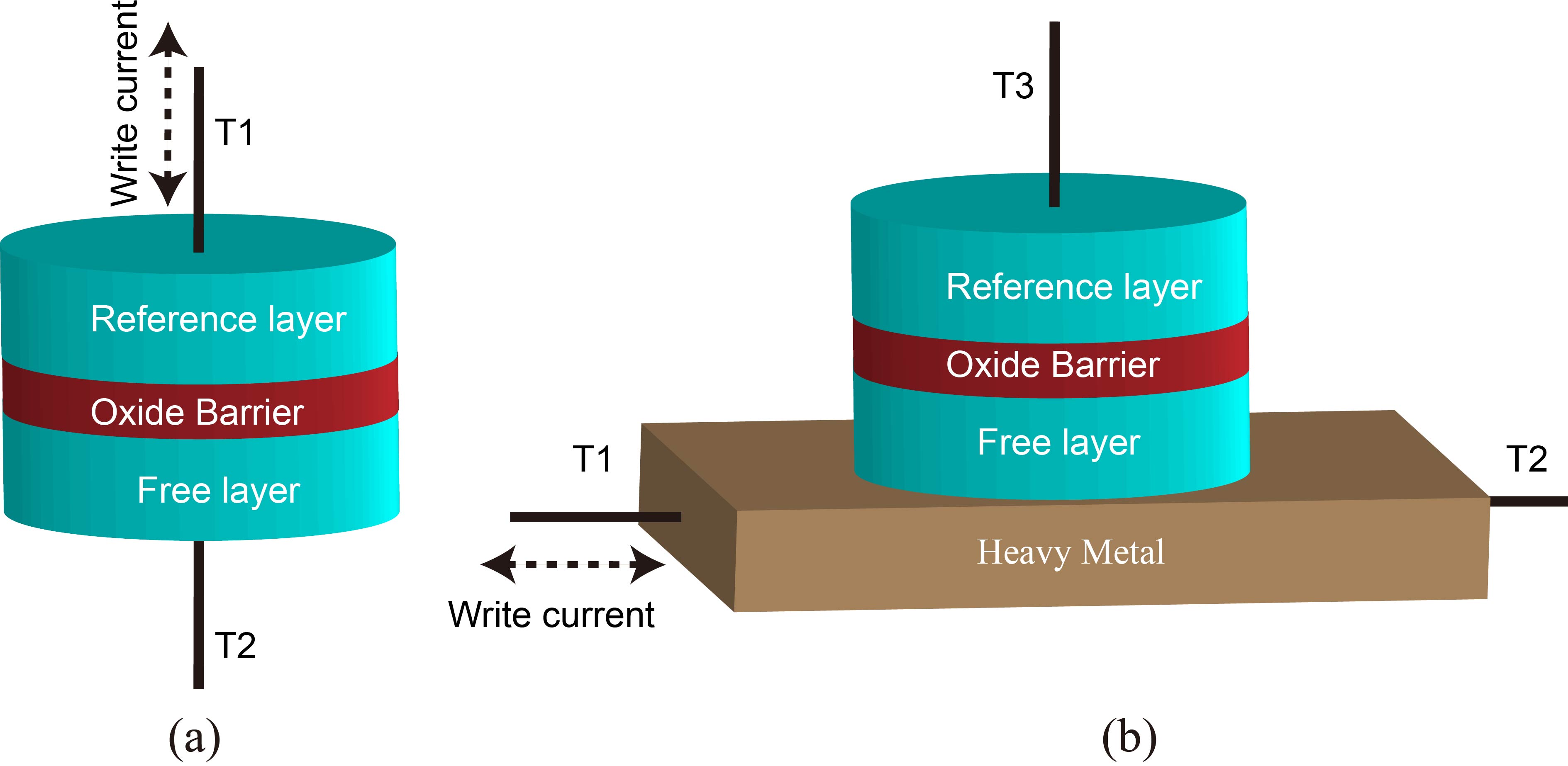}}
\caption{Illustration of the memory cells and the write strategies of STT-MRAM (a) and SOT-MRAM (b).}
\label{MRAMbit}
\end{figure}

SOT-MRAM utilizes the spin-orbit torques to write the memory cell, overcoming the drawbacks of Spin Transfer Torque-MRAM (STT-MRAM) in terms of high write latency, and large write energy dissipation\cite{wang2018high} \cite{chang2017prescott}. Fig.~\ref{MRAMbit} compares the similarity and difference between SOT-MRAM and STT-MRAM cells. Similarly, both types of MRAM cells store the bit value in a magnetic tunnerling junction (MTJ). The bit value "0" or "1" is read out electrically as high or low tunneling magneto-resistance, which is controlled by the antiparallel (AP) or parallel (P) alignment of magnetization in the free layer (FL) and the reference layer (RL). Although the write of MRAM bit is always fulfilled by controlling the magnetization direction of the FL, the mechanisms used are different between STT-MRAM and SOT-MRAM. In STT-MRAM, the write current passes through MTJ and the spin polarized current exerts notable STT to switch the FL magnetization\cite{wang2018high} \cite{chang2016evaluation}. Differently in SOT-MRAM, SOTs are generated by transversing write current though an additional heavy metal layer (HML) to switch the magnetization in the adjunct FL. As a result, SOT-MRAM does not suffer from the asymmetric of write latency between "AP $\rightarrow$ P" and "P $\rightarrow$ AP" in STT-MRAM, speeding up the write procedure. Moreover, the energy efficiency of write is fundamentally higher in SOT-MRAM. That's because each electron can be reused multiple times to exert SOTs after bounced back from HML and FL interface, while it can be used once at most in STT.

\begin{figure}[htbp]
\centering
\centerline{\includegraphics[width=3 in] {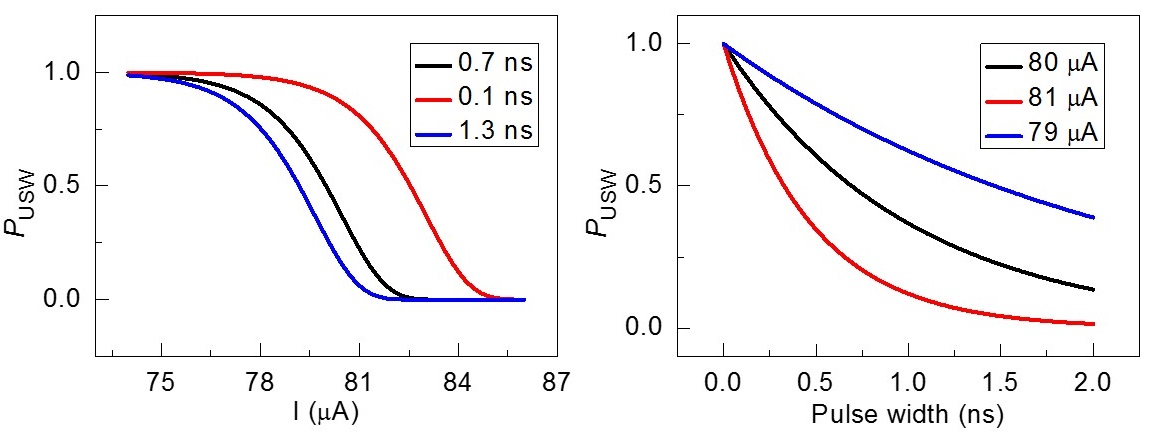}}
\caption{The probability of a MRAM bit remaining unswitched under different pulse duration and strength of the applied electrical current}
\label{Probability}
\end{figure}

We harness the stochastic behavior within the memory write of SOT-MRAM to perform SC. The probability $P_{usw}$ of MRAM bit remains \textbf{not} switched under the appliance of electrical current $I$ is described \cite{seki2011switching}
\begin{equation}
P_{usw}=exp(-\tau exp(-\Delta (1-I/I_c))).
\end{equation}
Here, $\tau$ denotes the pulse duration of the applied $I$ in nanosecond, $\Delta$ represents the thermal stability parameter of the MTJ, and $I_c$ is the critical current strength required to switch the FL magnetization. Fig.~\ref{Probability} plots the $P_{usw}$ as functions of $I$ and $\tau$, with $\Delta=60.9$ and $I_c=80\mu A$ estimated from previous micromagnetic simulations on SOT driven magnetization dynamics\cite{chang2017prescott}. By finely controlling the parameters in the write of SOT-MRAM, each memory bit can serve as a stochastic bit generator with the desired probability of holding either "0" or "1". Utilizing this feature of SOT-MRAM, the large amount of stochastic bits in SC can be generated in parallel and in-situ stored in memory with a simple write operation. 

\section{Data conversion and hardware design}
To implement the MUL operation with the stochastic switching of MRAM bit, the binary operands have to be translated into certain parameter of the write voltage pulse. The flow of our proposed sequential data conversion can be summarized as: 
\begin{equation}
\begin{split}
&X(binary), Y(binary) \rightarrow ln(X)(binary), ln(Y)(binary)\\
&\rightarrow ln(X)(time), ln(Y)(time) \rightarrow X*Y (stoch.) \\
&\rightarrow X*Y (Binary)
\end{split}
\end{equation}
In this section, we will introduce them and the related hardware design step by step.

\subsection{Binary numbers to logarithmic timing signals}
\begin{figure}[htbp]
\centering
\centerline{\includegraphics[width=3 in] {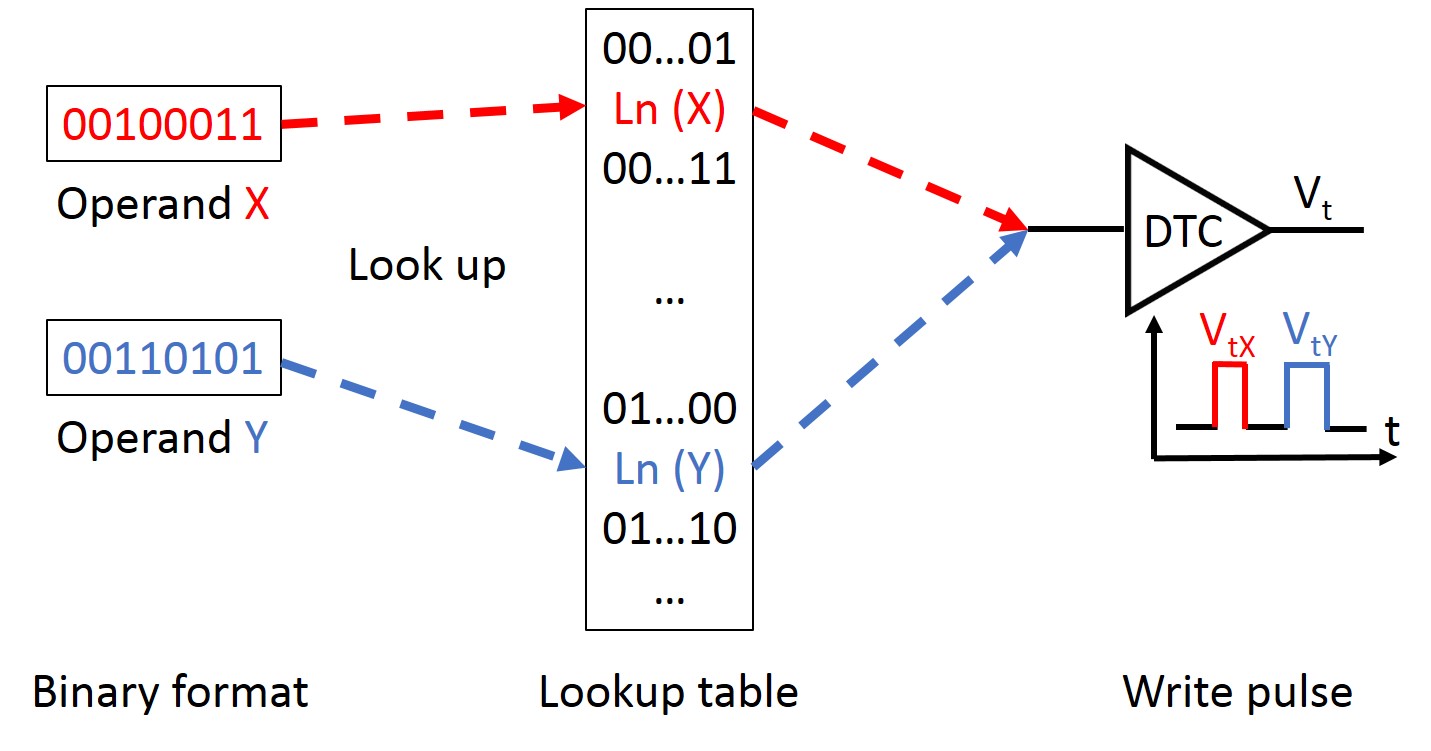}}
\caption{Data conversion from binary to logarithmic timing signals}
\label{LUT}
\end{figure}

We first perform logarithmic operation on the digital numbers stored in memory, i.e. $X(binary), Y(binary) \rightarrow ln(X)(binary), ln(Y)(binary)$. The multiple bits of the operand $X$ are read out by sensing amplifiers (SAs) and decoded to find their logarithmic values using a lookup table (LUT) (Fig.~\ref{LUT}). The LUT method is usually used in logarithm multiplication, and has been demonstrated to be fast and accurate\cite{nandan201865}. This conversion step is necessary, since an exponential operation is inherently included in the following stochastic switching of the MRAM bit. 
   
Afterwards, we convert the $ln(X)$ to timing signals with a digital-to-time converter (DTC). The DTC outputs a voltage square pulse $V_{tX}$, where the pulse duration $\tau_{X}$ in Eq. 3 is proportional to the value of input $ln(X)$. The magnitude of the $V_{tX}$ pulse is normalized and fixed to drive SOT-MRAM bit in its non-deterministic switching region.  
\subsection{Logarithmic timing signals to stochastic bitstream}

\begin{figure}[htbp]
\centering
\centerline{\includegraphics[width=3 in] {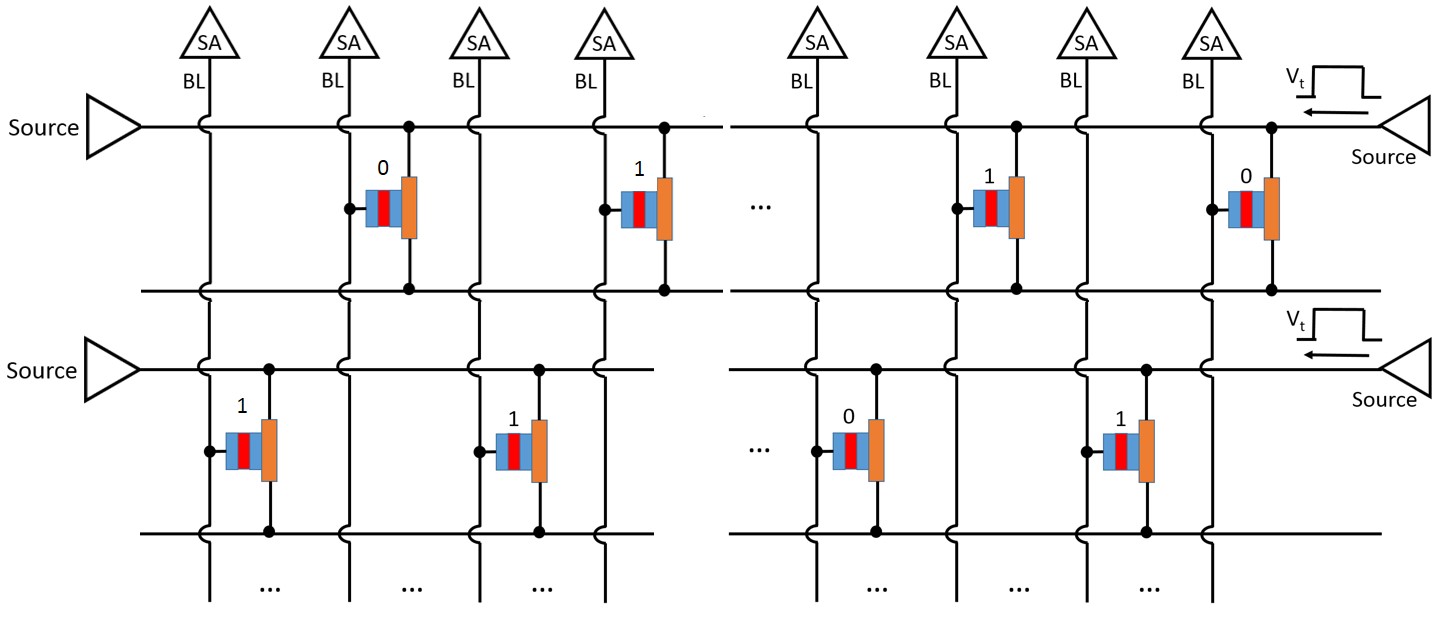}}
\caption{Schematics of the cross-point SOT-MRAM array, with stochastic bits in-situ stored under a pulsed voltage}
\label{Crosspoint}
\end{figure}

The write voltage pulse $V_{tX}$ is subsequently applied onto the source lines (SLs) of multiple rows of SOT-MRAM bits, and drives their stochastic switching behaviors. The entire row of MRAM array can be written simultaneously with a cross-point design (Fig.~\ref{Crosspoint})\cite{chang2017prescott}. The MTJs in a row share a set of driving transistors, and are directly linked to the BLs and SLs without additional transistors for individual bit. As a result, minimal area and energy overhead are introduced to enable such simultaneous write.

\begin{figure}[htbp]
\centering
\centerline{\includegraphics[width=3 in] {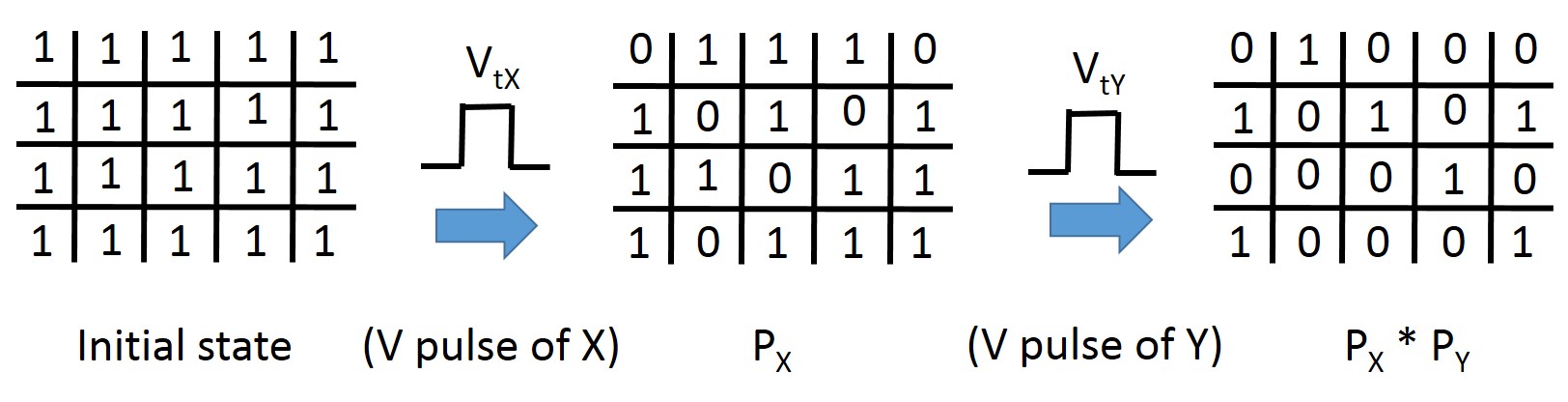}}
\caption{Illustration for the SC-based MUL in SOT-MRAM}
\label{SCexample}
\end{figure}

Fig.~\ref{SCexample} shows how the SC-based MUL is performed. \textbf{Initialization}: a preset operation is required to initialize all the bits to "1" with reversed current $I_c$. \textbf{Input first operand}: the converted write voltage pulse $V_{tX}$ is input onto the MRAM array, resulting in partial switching of the bits. The probability of remaining "1"s equals to $P_X$, where $P_X$ is proportional to the value of operand $X$. \textbf{MUL with the second operand input}: The MUL operation is performed by inputing a subsequent voltage pulse $V_{tY}$ (similarly converted from operand $Y$) onto the same MRAM array. As a result, the remaining "1"s survive from not switched by neither pulse $V_{tX}$ nor $V_{tY}$, and they are distributed among the MRAM arrays with a probability equaling $P_X*P_Y$ (proportional to $X*Y$).

\subsection{Stochastic bits to Binary numbers}
At last, we perform bit counting to convert the outcome from the stochastic representation to its binary format. Either approximate pop-count (APC) \cite{kim2015approximate} or PIM-based ADD operations \cite{li2017drisa} can be employed to bit counting. APC method can be performed with one clock cycle, but introduces much area overhead. Alternatively, PIM-based ADD is area-efficient, but takes many clock cycles to perform the pop-count.

\begin{figure}[htbp]
\centering
\centerline{\includegraphics[width=3.5 in] {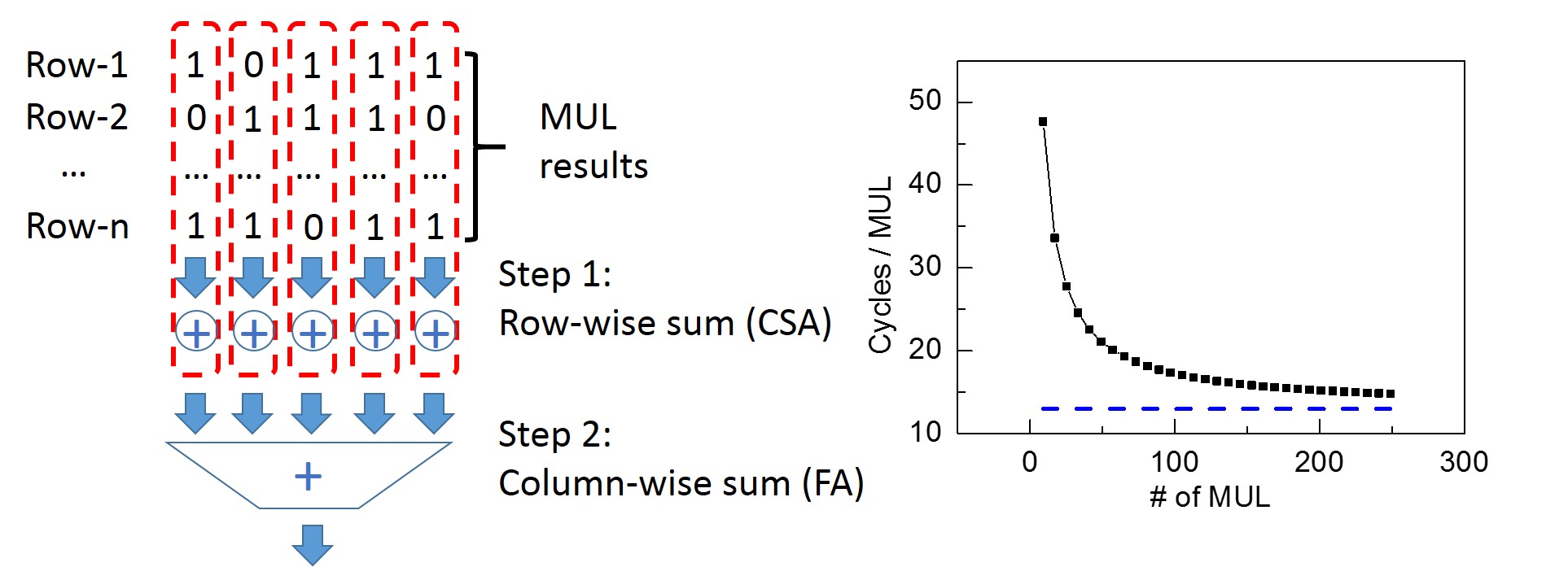}}
\caption{The two steps of the PIM-based pop-count strategy.}
\label{Popcount}
\end{figure}

Specifically, we can accelerate the PIM-based pop-count for the vectored multiply-and-accumulate (MAC) in NN. Fig.~\ref{Popcount} shows the two-step strategy, where the sum is performed after several MULs have been done. In the first step, we perform row-wise sum with a carry-save addition (CSA). Then in the second step, the intermediate sum results undergo a column-wise additions with full adder (FA). Our motivation here is to lessen the usage of FA for column-wised addition, since it takes more clock cycles than the lock step bitwise operations of CSA. As shown in Fig.~\ref{Popcount}, the delay from FA can be averaged out, and the pop-count related cycle count converges to that of CSA after many MULs. 

\subsection{Put them all together}
After putting all the pieces together, we point out strategies to further improve the performance and accuracy, and explain certain considerations in the design.

The sequential flow of data conversion can be separated and pipelined to improve the throughput and performance. For example, the LUT operation on the second operand can be performed simultaneously with the stochastic memory write for the first operand. Moreover, the bit counting can work in parallel with MUL operations for NN applications. There is no need for the relative slow pop-count to start until all the fast MULs between $w_i$ and $x_i$ have been finished in the computation of $\sum_iw_ix_i$. Furthermore, one could pre-convert certain frequently used data (e.g. weight $w_i$ in NN) into stochastic bits, which can be stored non-volatilely in MRAM arrays. Once other multipliers (e.g. inputs $x_i$) come, their converted timing signals can be directly input onto the corresponding MRAM arrays to perform MUL operations.  

There are several normalization units in the circuits that can be used to fine tune the accuracy and performance. For example, the pulse duration of $V_t$ can be scaled to a range where $P_{usw}\approx 0.5$. Through such scaling, the switching voltage pulse can not be longer than the usual time required to switch MRAM bit, avoiding unnecessary slowdown in computing. Moreover, the bitstream can be tuned neither sparse nor dense to guarantee the accuracy of MUL, so that more bits are effectively involved in SC. This is fundamentally similar to the improved classification accuracy of NN with more neurons involved.

Multiple rows can be simultaneously activated and wrote to generate more stochastic bits in parallel. This situation happens when performing MUL on operands with more bits. In the cross-point MRAM design, we limit the number of memory cells in each row due to the concern of IR drop. The MTJs farther away from the driving transistors in the row would suffer from a lower switching voltage\cite{liang2010cross}, and would likely undergo stochastic switching with undesired and incorrect probability.

Finally, we note that the pulse duration $\tau$ is used here for computing, instead of the magnitude of switching voltage pulse $V_t$ (equivalent to $I$ in Eq. 3). That's because the usage of the magnitude $I$ requires more complicated circuits design for data conversion, owing to the complex dependence of $P_{usw}$ on $I$. In addition, the two inputs $I_X$ and $I_Y$ has to be input simultaneously onto the MRAM arrays. This is not friendly to the pipeline strategies mentioned above, but will introduce large area overhead onto the driven transistors to enable higher current write instead. 

\section{Monte Carlo simulations}
To estimate the accuracy and its dependence on hardware variance from statistics, we performe the Monte Carlo simulations on the stochastic switching of MRAM bits. In the following, $nbit$ denotes the number of stochastic bits per MUL, $P$ represents the probability $bitcount(B_i=1, i=1:nbit)/nbit$ that the bit remains not switched under certain input voltage pulse. For one MUL operation, we test the proposed SC with 1000 iterations and make statistics on the results among iterations.

\subsection{Accuracy}
\begin{figure}[htbp]
\centering
\centerline{\includegraphics[width=3 in] {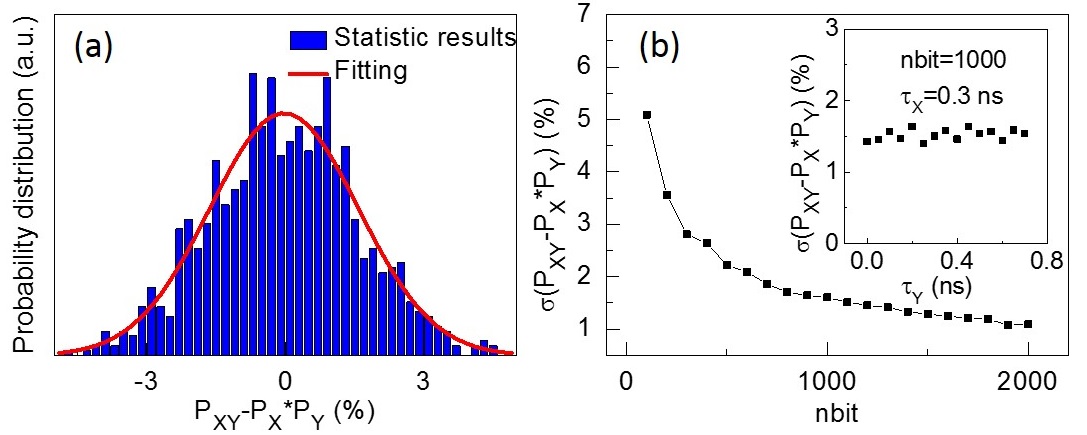}}
\caption{(a) The probability distribution of the deviation from theoretical calculated MUL among 1000 iterations. (b) The MUL uncertainty as a function of stochastic bits number and input values.}
\label{Statadd}
\end{figure}

Fig.~\ref{Statadd}(a) shows the distribution of the error $P_{XY}-P_X*P_Y$ among the 1000 iterations, where the the probability $P_{XY}$ is stochastically computed (with $\tau_{X}=0.3 ns, \tau_{Y}=0.4 ns$) and $P_X, P_Y$  are theoretically calculated from the two operands. The error distribution is centered to zero, indicating that there is no intrinsic bias in the SC arithmetic. The distribution can be well fitted with a Gaussian function (red line), with the standard deviation $\sigma \approx 1.6\%$. This indicates that the MUL is with about $3.2\%$ uncertainty for $nbit=1000$.

We further investigate the dependence of $\sigma$ on the inputs $\tau_Y$ and the number of stochastic bits $nbit$. As shown in Fig.~\ref{Statadd}(b), $\sigma$ is almost independent on the inputs $\tau_Y$, but decreases with larger $nbit$. Therefore, we can improve the accuracy of SC by using more MRAM bits, despite that the improvement becomes more gradual with larger $nbit$. 

\subsection{The impact of hardware variance}
We also investigate the impact of hardware variance on the accuracy of MUL operation, by introducing random fluctuations on the devices' parameters in Monte Carlo simulation. 

\begin{figure}[htbp]
\centering
\centerline{\includegraphics[width=3 in] {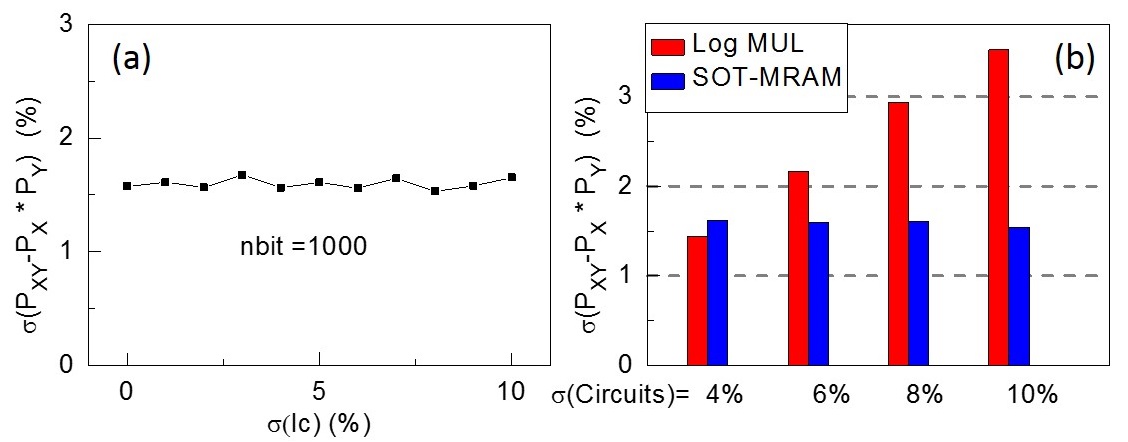}}
\caption{The dependence of MUL uncertainty on the variance of critical current among SOT-MRAM bits. The MUL uncertainties as functions of circuit variance in our SC+PIM design and logarithm multiplication}
\label{Faulttorr}
\end{figure}

The critical currents of MRAM bits $I_c$ may be slightly varied, since the many MRAM bits can not be manufactured identically and they may also experience different thermal fluctuations when in use\cite{an2016current}. Therefore, we introduce 0\% to 10\% random fluctuations $\sigma(I_c)$ on the $I_c$. As shown in Fig.~\ref{Faulttorr}(a), the accuracy of SC remains almost unchanged under different strength of fluctuations. 

We also compare the fault tolerance of our design with that of logarithm multiplication. To implement logarithm multiplication\cite{nandan201865}, we replace the DTC and SOT-MRAMs with an antilogarithm amplifier. Then we introduce 4\% to 10\% random fluctuations $\sigma(Circuits)$ on DTC and antilogarithm respectively for the two cases. As shown in Fig.~\ref{Faulttorr}(b), the accuracy of our SC+PIM design remains almost unchanged, while logarithm multiplication suffers from severe degradation in accuracy with stronger fluctuations. 

\section{Evaluation}
In this section, we evaluate the performance, power and area overhead of the proposed SC+PIM design, and compare them with that of other approaches using either SC or PIM. 

\subsection{Experimental setup}
We adopt the cross-point design of SOT-MRAM arrays similar to PRESCOTT\cite{chang2017prescott}, to enable the parallel memory write. The low-power DTC generates voltage pulses with 22 ps time resolution and occupies $75\mu m*25\mu m$ in area\cite{wang2015digital}. For the APC, we design one-cycle fully parallel circuit synthesized with 45nm FreePDK\cite{stine2009freepdk}, integrating parameters from\cite{kim2015approximate}. Our evaluation is based on the multiplication between two 10-bit operands that represented by $2^{10}$ stochastic bits. 

In the following, different configurations have been compared: \textbf{SC+PIM (with APC)} denotes our SC+PIM design with pop-count conducted by APC. \textbf{SC+PIM (with CSA)} is our SC+PIM design with pop-count performed with CSA+FA. Specially, the evaluation is averaged onto each MUL for the situation of performing 100 MULs in a MAC. \textbf{SC} represents the usage of a built multiplier with the state-of-the-art SNG \cite{kim2016energy} and popcount with APC. \textbf{PIM} is the situation that we only use in-memory Boolean logic operations to implement MUL.  

\subsection{Performance}
\begin{figure}[htbp]
\centering
\centerline{\includegraphics[width=3.5 in] {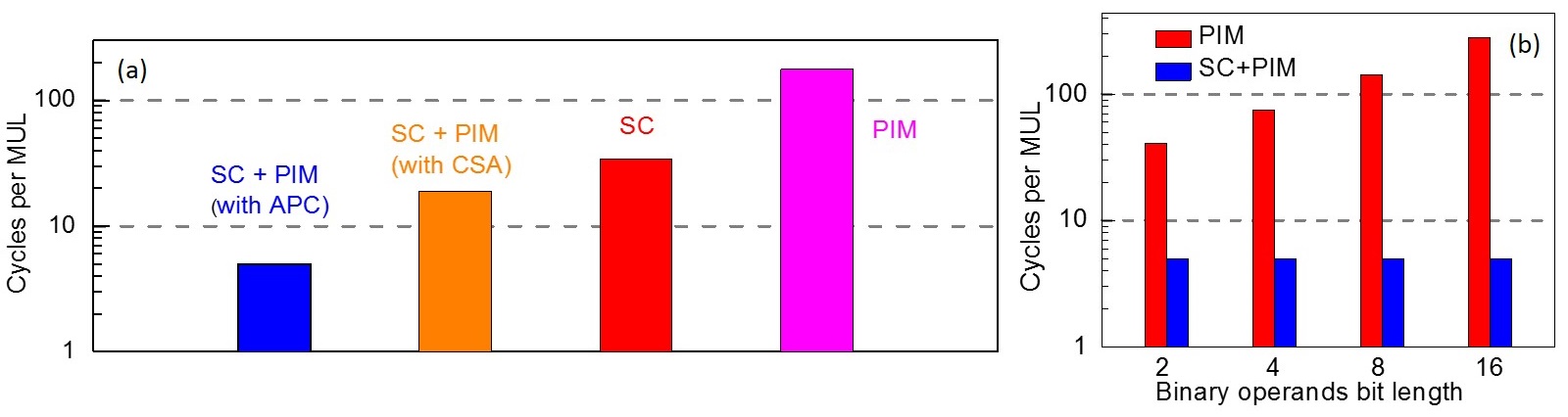}}
\caption{(a) The cycle count to implement each MUL with different approaches. (b) The MUL latency as a function of multipliers' bit length with SC+PIM and PIM approaches}
\label{Performance}
\end{figure}

Fig.~\ref{Performance}(a) compares the cycle count used to perform each MUL operation with different designs. Evidently, our SC+PIM approach outperforms prior approaches using either SC or PIM. The boost of performance in our design benefits from the parallel generation of stochastic bits. In contrast, prior SC approaches requires additional cycles to generate stochastic bitstreams or to shuffle the existing pseudo-stochastic or deterministic bitstreams \cite{kim2016energy}. 

In addition, we investigate the dependence of MUL cycle count on the operands' bit length as shown in Fig.~\ref{Performance}(b). The cycle count remains unchanged in our SC+PIM design, since different amount of stochastic bits ($2^n$ for n-bit operand) can be generated in parallel. As a comparison, the cycle count required for MUL increases exponentially for the operands' bits length in prior PIM design. Therefore, the speedup of SC+PIM over PIM becomes more attractive for MUL between operands with more bits. 

\subsection{Energy consumption}
\begin{figure}[htbp]
\centering
\centerline{\includegraphics[width=3 in] {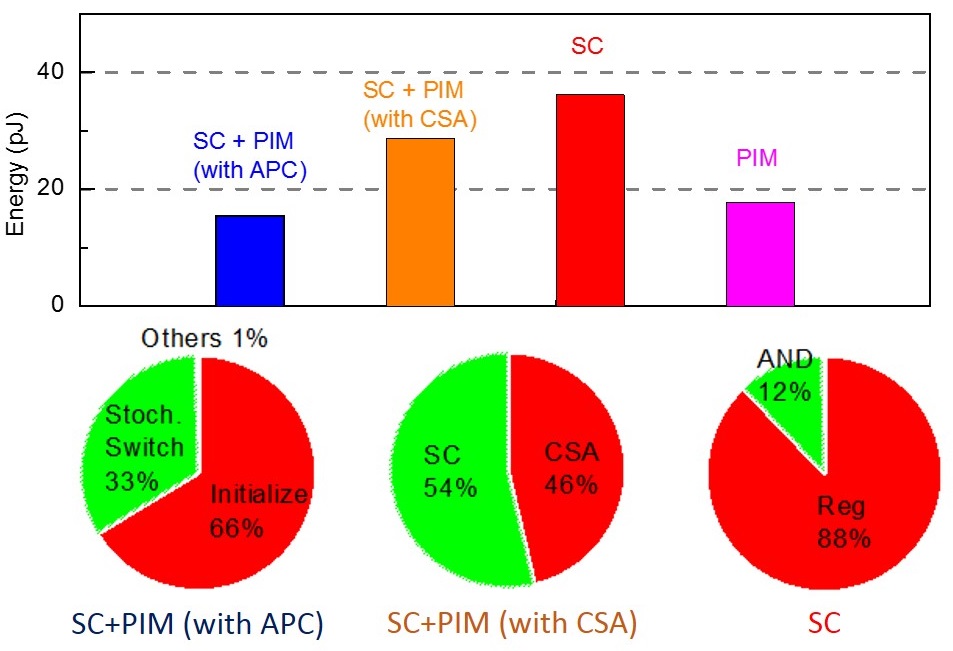}}
\caption{The energy consumption for each MUL with different approaches and their breakdown.}
\label{Energy}
\end{figure}

Our SC+PIM design consumes 58\% less energy compared with the SC method (Fig.~\ref{Energy}), thanks to the low write energy of SOT-MRAM\cite{jabeur2014spin}. In our design, most energy is spent through memory write, such as in the generation/computing of stochastic bits and the pop-count with bitwise addition (CSA). The situation is similar to prior SC approaches, where 88\% of the energy is consumed in data buffering related operations. 

As shown by the breakdown of the energy consumption in Fig.~\ref{Energy}, the initialization step costs more energy than the following steps performing SC for MUL. That's because a write voltage pulse with a higher magnitude and a longer pulse duration needs to be applied to guarantee the initialization. Afterwards, the memory bits are mainly driven in a non-deterministic switching region which consumes less energy. 

\subsection{Area overhead}
\begin{figure}[htbp]
\centering
\centerline{\includegraphics[width=3 in] {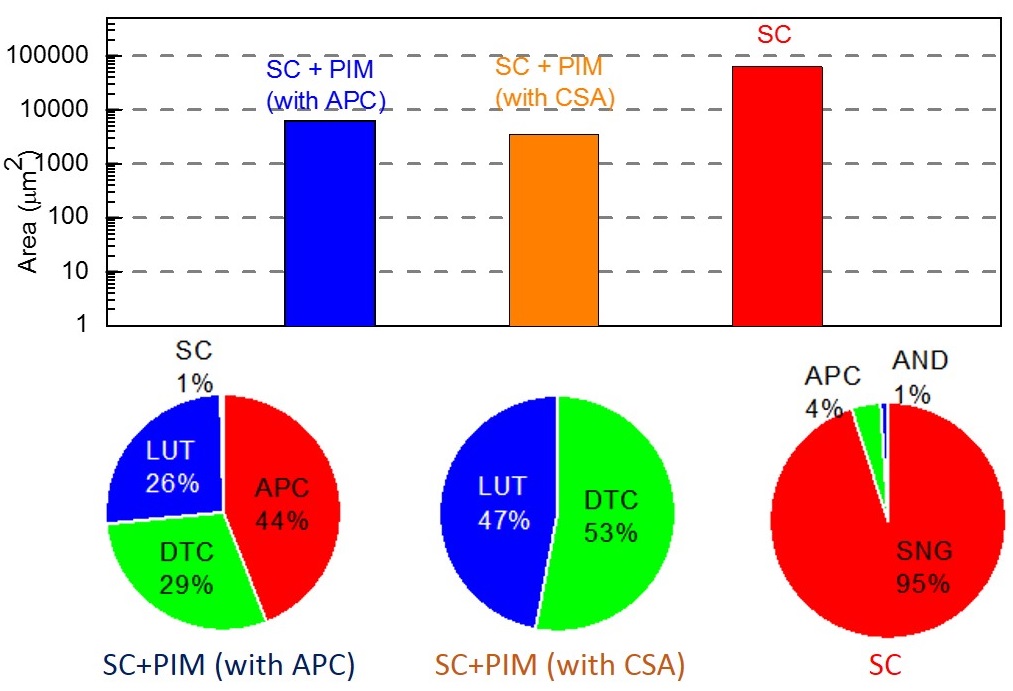}}
\caption{The area overhead in different approaches with their breakdown. }
\label{Area}
\end{figure}

The area overhead of different designs is compared in Fig.~\ref{Area}. The area overhead is smaller by about one order of magnitude for our SC+PIM design than conventional SC. The improvement originates from the removal of the additional circuits for SNG, which occupies 95\% of the area in the conventional SC approach. 

As shown by the breakdown of area overhead in Fig.~\ref{Area}, the memory space required for the LUT table is comparable to the DTC and APC  in our design, for the case of 10-bit multiplication. The LUT table size will shrink for regular 8-bit multiplication, since it depends exponentially on the bit length of the operands. 

\section{Conclusion}
In this paper, we propose a new SC design to perform MUL with in-memory operations. The stochastic random generation and AND operation in conventional SC are implemented by the simple write operations onto the SOT-MRAM. Such design is enabled by converting the binary multipliers to the varied pulse duration of the write voltage for SOT-MRAM. Consequently, the stochastic bits for the MUL outcome are in-situ stored. Two strategies of pop-count (APC or PIM-based ADD) have been proposed to convert the MUL result back to its binary format, offering flexibility to further trade off performance with area. Our approach improves the performance to compute MUL with PIM, in synergy with the mitigation of area overhead for supporting circuits of SC. 

\bibliographystyle{IEEEtran}


\end{document}